\documentclass[final,cite]{cimento}

\usepackage{amsmath,amsfonts,amssymb,bm}
\usepackage[all]{xy}
\usepackage[final]{graphicx}

\title{Optically induced coherence effects in `artificial atoms and molecules'}

\author{Ulrich Hohenester\from{graz}
        \thanks{Electronic mail: ulrich.hohenester@uni-graz.at}, 
        Filippo Troiani\from{modena} \atque
	Elisa Molinari\from{modena}}
	
\instlist{
  \inst{graz}
    Institut f\"ur Theoretische Physik,
    Karl--Franzens--Universit\"at Graz, Universit\"atsplatz 5,
    8010 Graz, Austria
  \inst{modena}
    Istituto Nazionale per la Fisica della Materia (INFM-$S^3$) and
    Dipartimento di Fisica,\\
    Universit\`a degli Studi di Modena e Reggio Emilia,
    Via Campi 213a, 41100 Modena, Italy}
    
\PACSes{
  \PACSit{78.67.H}{Quantum dots (optical properties)}
  \PACSit{78.47.+p}{Time-resolved optical spectroscopies in condensed matter}
  \PACSit{42.50.M}{Self-induced transparency}
}

\checkin{April 22, 2002}{---}

\newcommand*{\E}[0]{\mathcal{E}}
\renewcommand*{\P}[0]{\mathcal{P}}
\newcommand*{\J}[0]{{\bm{\mathcal{J}}}}
\newcommand*{\stirap}[0]{{\sc stirap}}
\newcommand*{\sit}[0]{{\sc sit}}
  
\begin{document}

\maketitle
  
\begin{abstract}

Using a master-equation approach for the description of coherent and incoherent dynamics in `artificial atoms and molecules', we present a theoretical analysis of situations where intense laser fields lead to pronounced renormalizations of carrier states. Such enhanced light-matter interactions allow for the solid-state implementation of effects hitherto only observable in atomic systems. Two prototypical examples will be presented: first, we show how two intense laser pulses can be exploited for a robust and high-fidelity population transfer of carrier states in coupled quantum dots; second, we discuss the possibility of observing self-induced transparency in a sample of inhomogenously broadened quantum dots, a phenomenon where intense laser pulses propagate without suffering significant losses.

\end{abstract}

\section{Introduction}

{\em Quantum Optics}\/ exploits the laws of quantum mechanics to optically control atomic systems with highest possible precision \cite{mandel:95,scully:97}: recent years have seen spectacular examples of such light-matter manipulations, {\it e.g.},\/ Bose-Einstein condensation or freezing of light \cite{nature:02}, as well as the emerging fields of quantum computation and quantum communication \cite{bouwmeester:00}. Evidently, the driving force behind all these efforts is to explore the consequences of quantum mechanics in real physical systems---possibly up to their most far reaching limits.

This tremendous success of {\em Quantum Optics}\/ also initiated great stimulus in the field of solid-state physics. In particular with semiconductor quantum dots \cite{woggon:97,hawrylak:98,bimberg:98}, or {\em `artificial atoms'}\/ as they are sometimes called because of their atomic-like carrier states, one now has a system at hand which resembles many of the atomic properties whilst offering at the same time all the flexibility of semiconductor nanostructures: such remarkable features suggest the implementation of quantum-optical schemes in this novel class of materials.

Indeed, this synthesis of {\em Quantum Optics}\/ and low-dimensional semiconductor physics formed the primary research topic of Giovanna Panzarini's activities. In the last years her main interest was focused on the identification of quantum-optical schemes with a successful history in atom physics and a promising future for quantum-dot implementations. Two of these proposals, which Giovanna developed in collaboration with us, will be reviewed here. We have organized our paper as follows: in sect. \ref{sec:intro-dot} we briefly summarize our present understanding of carrier states in quantum dots; based on this discussion, we set out in sect.~\ref{sec:theory} to develop our theoretical framework; sects. \ref{sec:stirap} and \ref{sec:sit} present results of our calculations where we analyze two prototypical examples of strong radiation-matter interaction; finally, we draw in sect.~\ref{sec:conclusion} some conclusions and present an outlook to future developments.

\subsection{Characteristics of artificial atoms and molecules}
\label{sec:intro-dot}

Quantum dots consist of a small island of lower-bandgap material embedded in a solid-state matrix of higher-bandgap material; proper choice of the material and dot parameters thus can give rise to the confinement of a few carrier states within this lower-bandgap region, resulting in discrete, atomic-like spectra and strongly enhanced lifetimes. Without aiming at an exhaustive discussion, in the following we outline our present understanding of carrier dynamics in quantum dots, which will guide us in developing our theoretical description scheme in sect.~\ref{sec:theory}. More specifically, we shall address few-particle states and Coulomb renormalizations, dephasing and relaxation, and interdot coupling and tailoring of optical properties.

In contradistinction to `natural' atoms, the confinement potential of artificial atoms is not identical because of dot-size and material fluctuations inherent to any growth procedure \cite{woggon:97,bimberg:98}. As consequence, all optical experiments suffer from large inhomogeneous line broadenings, which spoil the observation of finestructure splittings characteristic for zero-dimensional systems. A major advancement in the field has come from different types of local optical experiments, that allow the investigation of individual quantum dots thus avoiding inhomogeneous broadening \cite{zrenner:00,simserides.prb:00}: indeed, such {\em single-dot spectroscopy}\/ revealed a surprisingly rich fine-structure of the optical spectra for both multi-excitons \cite{zrenner:00} and multi-charged excitons \cite{warburton:00,hartmann.prl:00,findeis.prb:01}. For our present purpose, the essential characteristics of these findings can be summarized as follows: first, because of the strong confinement quantum dots can host various electron-hole complexes which would be unstable in semiconductors of higher dimensionality; second, whenever one or more carriers are put into a quantum dot the optical spectra change because of the resulting additional Coulomb interactions \cite{bayer:00a,hohenester.pss-b:00}. As consequence, each quantum-dot spectrum uniquely reflects its electron-hole configuration. Furthermore, because of dot-size fluctuations these spectral fingerprints vary from dot to dot, and thus allow to spectroscopically address specific quantum dots.

In the early days of dot spectroscopy there was an intense debate whether the phase-space restrictions imposed by the strong quantum confinement would inhibit phonon-induced carrier relaxation and dephasing of optically excited carrier states, and led to the prediction of the so-called ``phonon bottleneck'' effect. However, experimental evidence indicates that whenever an electron-hole complex can relax to a state of lower energy by phonon emission, the corresponding transition occurs on a relatively short timescale, and no clear evidence for a phonon bottleneck was reported. It should be emphasized, however, that the nature of phonon scatterings in quantum dots can be substantially different as compared to semiconductors of higher dimensionality: in particular if the energy difference between two states does not match the phonon energy, additional scattering and dephasing channels, such as polaron-mediated relaxation \cite{verzelen:00} or multi-phonon processes \cite{uskov:00}, are required. Generally speaking, for a given electron-hole configuration long dephasing and lifetimes can only be expected for the respective state of lowest energy, whereas ``excited'' states relax much faster. Recently, Borri \etal\ \cite{borri:01} extracted from low-temperature photon-echo experiments exciton lifetimes that were indeed solely governed by the radiative exciton decay.

Finally, we briefly discuss the possibility of tailoring dot and environment coupling. In most growth procedures it is relatively simple to produce vertically coupled quantum dots \cite{schedelbeck:97,bayer:01a}, which, in analogy to molecules, are sometimes referred to as {\em `artificial molecules'}.\/ These structures have recently attracted increasing interest both in view of basic research \cite{troiani.prb:02} as well as of possible quantum computation applications \cite{troiani.prb:00,biolatti:00}. Additionally, semiconductor quantum dots can be combined with other semiconductor-based optical components such as microcavities or photonic crystals, thus allowing for a highly flexible control of environment interactions.

\section{Theory}\label{sec:theory}

\subsection{Hamiltonian}

In the following we shall consider the problem of carriers confined in a quantum dot under the action of external laser fields. The carriers will experience their mutual Coulomb interactions as well as the interaction with phonons and other environmental degrees of freedom. It turns out to be convenient to describe the problem by the Hamiltonian:

\begin{equation}\label{eq:ham}
  H=H_o-\bm\E\hat{\bm P}+H_R+V
\end{equation}

\noindent where $H_o$ accounts for the carrier states; the light-matter interaction $-\bm\E\hat{\bm P}$ is described in the dipole approximation \cite{haug:93}, with $\bm\E$ the electric field of the laser and $\hat{\bm P}$ the interband polarization operator; finally, $H_R$ describes the environmental ({\em ``reservoir''}) degrees of freedom, which are coupled through $V$ to the carriers in the dot.

\subsubsection{Few-particle states}

From our introductory discussion it has become clear that Coulomb renormalization effects are of crucial importance for the description of few-particle states in semiconductor quantum dots. In contrast to higher-dimensional semiconductors, mutual Coulomb interactions among quantum dot carriers do not result in scattering and dephasing but solely give rise to energy renormalizations. Consequently, Coulomb correlation effects in dots are conveniently accounted for within a first-principles manner, {\it e.g.}, using the framework of configuration interactions \cite{rontani.ssc:01}. With $x$ labeling the few-particle states and $E_x$ the corresponding energies, the Hamiltonian accounting for the carrier states is of the form:

\begin{equation}\label{eq:h0}
  H_o=\sum_x E_x |x\rangle\langle x|.
\end{equation}

\begin{figure}
\[\begin{xy}0;<12mm,0mm>:
  (-1,2.6)*{\bm{(a)}},
  (0,-0.3);(1,-0.3)*{}**@{-},	
  (2,0);(3,0)*{}**@{-},
  (1,2);(2,2)*{}**@{-},  
  (0.5,-0.6)*{|1\rangle},
  (2.5,-0.3)*{|2\rangle},
  (1.5, 2.3)*{|3\rangle},
  (0.2,1.2)*{\Omega_p,\omega_p},
  (2.7,1.2)*{\Omega_s,\omega_s},
  (3.1,-0.15)*{\delta},
  (3,-0.3);(2.9,-0.3)*{}**@{-},
  (5,2.6)*{\bm{(b)}},
  (7,-0.3);(8,-0.3)*{}**@{-},
  (6.2,1.1);(7.2,1.1)*{}**@{-},
  (7.6,1.1);(8.6,1.1)*{}**@{-},
  (7,2.3);(8,2.3)*{}**@{-},
  (7.5,-0.6)*{|0\rangle},
  (5.8,1.1)*{|X_o^+\rangle},  
  (9.0,1.1)*{|X_o^-\rangle},
  (7.5,2.6)*{|B_o\rangle},
  (6.5,0.4)*{\sigma_+},
  (8.4,0.4)*{\sigma_-},
  (6.5,1.8)*{\sigma_-},
  (8.4,1.8)*{\sigma_+},  
  \ar@{<=>}(0.5,-0.2);(1.2,1.9)
  \ar@{<~} (0.7,-0.2);(1.4,1.8)
  \ar@{<=>}(2.5,0.1);(1.8,1.9)
  \ar@{<~} (2.2,0.1);(1.6,1.8)
  \ar@{.}(2.95,-0.28);(2.95,0)
  \ar@{<=>}(7.2,-0.2);(6.6,1.0)
  \ar@{<~} (7.4,-0.2);(6.8,0.9)
  \ar@{<=>}(7.8,-0.2);(8.2,1.0)
  \ar@{<~} (7.6,-0.2);(8.0,0.9)
  \ar@{<=>}(7.8,2.2);(8.2,1.2)
  \ar@{~>} (7.6,2.1);(8.0,1.2)  
  \ar@{<=>}(7.2,2.2);(6.6,1.2)
  \ar@{~>} (7.4,2.1);(6.8,1.2)  
\end{xy}\]
\caption{Two prototypical dot-level schemes. {\bf (a)} $\Lambda$-type scheme, {\it e.g.}\/ carrier states in coupled dots: $|1\rangle$ and $|2\rangle$ are long-lived states, whereas $|3\rangle$ is a short-lived state which is optically coupled to both $|1\rangle$ and $|2\rangle$ (for details see sec.~\ref{sec:stirap}); wiggled lines indicate spontaneous photon emission; ${\bf (b)}$ exciton states in a single dot: $|0\rangle$ is the vacuum state; $|X_o^\pm\rangle$ are the spin-degenerate single-exciton groundstates, and $|B_o\rangle$ is the biexciton groundstate whose energy is reduced as compared to $2\times E_{X_o}$ because of correlation effects \cite{zrenner:00,hohenester.pss-b:00}; optical selection rules for circularly polarized light, $\sigma_\pm$, apply as indicated in the figure.}\label{fig:scheme}
\end{figure}
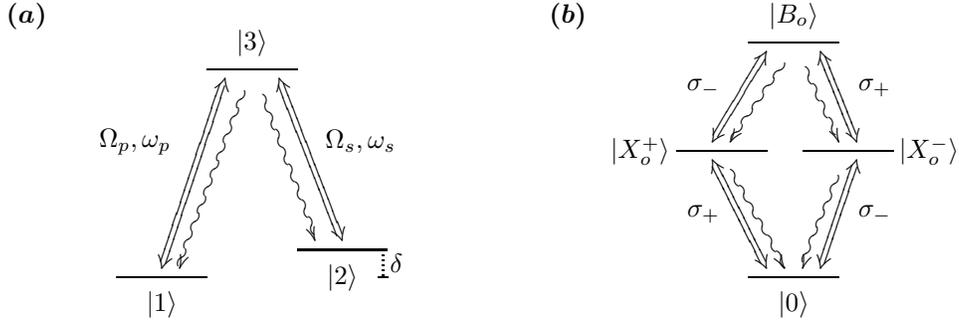

Note that, although the few-particle level scheme $E_x$ can be fairly complicated and dependent on a number of decisive material and dot parameters, in the simulation of the carrier time dynamics one often only needs a limited information about $E_x$. For instance, if the laser frequencies are tuned to the exciton groundstate, it completely suffices to know the energies of the exciton and biexciton groundstates together with the optical matrix elements connecting these states; these few parameters can then be inferred from experiment or calculated from theory (see, {\it e.g.},\/ fig. \ref{fig:scheme}). 

It thus is physical intuition together with the proper choice of the excitation scenario which allows to reduce a complicated few-particle problem, eq.~(\ref{eq:h0}), to a relatively simple few-level scheme. This situation is quite different as compared to the description of carrier dynamics in higher-dimensional semiconductors, where such a clear cut separation is not possible because of the scattering-type nature of carrier-carrier interactions; it is, however, completely similar to {\em Quantum Optics}\/ which relies on phenomenological level schemes, {\it e.g.}, $\Lambda$- or V-type schemes, with a few effective parameters---a highly successful approach despite the tremendously complicated nature of true atomic states.

\subsubsection{Optical excitation}

Suppose that a dot initially in its vacuum state, \idest\ no electrons and holes present, is subject to a laser excitation: the light-matter interaction will induce an {\em interband polarization},\/ which, in the language of semiconductor physics, can be described in terms of electron-hole pairs. Quite generally, the fast time dynamics introduced by the laser has to be treated with some care. To see that, we first note that the energy splitting between few-particle states $E_x$ with different numbers of electron-hole pairs $N_x$ is much larger than any energy splitting between states of the same $N_x$. Thus, we separate in eq.~(\ref{eq:h0}) these different energy scales {\em viz.}:

\begin{equation}
  H_o=\bar H_o+H_o'=
    \sum_x(N_x\omega_o)|x\rangle\langle x|+
    \sum_x\Delta_x|x\rangle\langle x|,
\end{equation}

\noindent where $\omega_o$ is an energy of the order of the bandgap ($\hbar=1$ throughout) and the detunings $\Delta_x=E_x-(N_x\omega_o)$ are small as compared to $\omega_o$. Correspondingly, we express the electric field of the laser as $\frac 1 2(\bm\E_oe^{-i\omega_ot}+c.c.)$, with the time-dependent envelope $\bm\E_o$. In what follows, we make use of the dipole and rotating-wave approximations: the first approximation states that light can only induce transitions between states differing by {\em one}\/ electron-hole pair; within the interaction representation according to $\bar H_o$, the interband dipole operator $\hat{\bm P}$ of eq.~(\ref{eq:h0}) can thus be split into terms with time dependencies of $e^{\pm i\omega_ot}$, respectively; in the latter approximation, we keep in $\bm\E\hat{\bm P}$ only terms where this fast oscillatory time dependence cancels out and neglect the remaining $e^{\pm i(2\omega_o)t}$ terms in the spirit of the random-phase approximation. Hence, the light-matter interaction reads:

\begin{equation}\label{eq:hop}
  H_{op}=-\mbox{$\frac 1 2$}\sum_{xx'} 
    \left(\bm\E_o^*e^{ i\omega_ot}\bm M_{x'x}+
          \bm\E_o  e^{-i\omega_ot}\bm M_{xx'}^*\right)
    |x'\rangle\langle x|,
\end{equation}

\noindent with $\bm M_{x'x}$ the optical dipole matrix elements \cite{hohenester.pss-b:00,haug:93}, where state $x'$ has one electron-hole pair less than $x$. Further below we shall find it convenient to introduce the (time-dependent) Rabi frequencies $\Omega_{x'x}=\bm\E_o^*\bm M_{x'x}$.

\subsubsection{Environment coupling}

Although in this paper we consider spontaneous emission of photons as the only source of dephasing and relaxation, we shall find it convenient to describe environment coupling in terms of a more generalized Caldeira--Leggett-type model \cite{leggett:87,hohenester.ssc:01}

\begin{equation}\label{eq:caldeira}
  H_R+V=\sum_\ell \omega_\ell a_\ell^\dagger a_\ell+
  i\sum_{xx',\ell}\left(g_{x'x}^\ell  a_\ell^\dagger-
                        g_{xx'}^{\ell\;*}a_\ell\right)
  |x'\rangle\langle x|,
\end{equation}

\noindent suited also for the description of, {\it e.g.},\/ phonon interactions. Here $\ell$ labels the environmental degrees of freedom with energy $\omega_\ell$; $a_\ell$ and $a_\ell^\dagger$ are the bosonic field operators; and $g_{x'x}^\ell$ are the matrix elements for transitions between the few-particle states $x$ and $x'$ through excitation of $\ell$. For the coupling to the propagating photon modes we obtain: $\ell=(\bm k\sigma)$, with $\bm k$ and $\sigma$ the photon wavevector and polarization, respectively; $\omega_{\bm k\sigma}=c'k$, with $c'$ the speed of light in the semiconductor; $g^{\bm k\sigma}_{x'x}=(2\pi\omega_{\bm k\sigma}/\kappa)^\frac 1 2 \hat{\bm e}_{\bm k\sigma}^* \bm M_{x'x}$, with $\kappa$ the semiconductor dielectric constant and $\hat{\bm e}_{\bm k\sigma}$ the unit vector of the light polarization.

\subsection{Dynamics}

After having specified our theoretical model in the previous section, we now set out to derive our central master equation; in so doing  we shall be guided by the different time scales introduced by the laser excitation (fast) and dephasing and relaxation (slow). We hence separate the {\em coherent}\/ and {\em incoherent}\/ time dynamics introduced by $H_{op}$ and $H_R+V$, respectively. As we aim at a description of the coherent and incoherent dynamics, we describe the quantum dot system through its density-matrix $\bm\rho$ \cite{scully:97,haug:93}, whose diagonal elements $\bm\rho_{xx}$ describe the occupation of the few-particle states $x$, and the off-diagonal terms $\bm\rho_{xx'}$ account for the coherence between states $x$ and $x'$.

\subsubsection{Coherent time evolution}

From eqs.~(\ref{eq:h0}) and (\ref{eq:hop}) and using the Liouville von-Neumann equation we obtain ---within the interaction representation according to $\bar H_o$--- for the coherent time dynamics $\dot{\bm\rho}=-i[\bm h_o+\bm h_{op},\bm\rho]$, where:

\begin{equation}\label{eq:coherent}
  (\bm h_o+\bm h_{op})_{xx'} = 
    \delta_{xx'}\Delta_x -
    \mbox{$\frac 1 2$} (\Omega_{xx'}+\Omega_{x'x}^*).
\end{equation}

\subsubsection{Incoherent time evolution}

In contrast to the coherent time dynamics, coupling to a reservoir, which acts as a heat bath, cannot be described exactly. Hence, we are forced to pursue from the beginning an approximative approach. Quite generally, we shall make the following assumptions: firstly, scatterings occur seldomly on a time scale characteristic for the laser-induced coherent dynamics---an assumption which certainly holds for low temperature and considering only the groundstates in the subspaces of given $N_x$; secondly, scatterings are fast processes which can be treated as almost instantaneous---a well-controlled approximation which imposes some restrictions on the spectral density of the reservoir and the dot-reservoir coupling strength \cite{scully:97,walls:95}; thirdly, in the description of scatterings the carrier dynamics is dominated by $H_o$, and $H_{op}$ only introduces minor corrections. In more technical terms, these assumptions guarantee that we can use the adiabatic and Markov approximations. Hence, the dot-reservoir interaction can be described in second order time-dependent perturbation theory as \cite{walls:95}:

\begin{equation}\label{eq:reservoir}
  \dot{\bm\rho_t}\Bigl|_R=-\int^t_{t_o} d\bar t\; \mbox{tr}_R
  [ V_t,[V_{\bar t},\bm\rho_R\otimes\bm\rho_t]],
\end{equation}

\noindent with $\bm\rho_R$ the density matrix of the reservoir and $\mbox{tr}_R$ denoting the trace over the environmental degrees of freedom (note that in eq.~(\ref{eq:reservoir}) we have used an interaction representation according to $H_o+H_R$). In the evaluation of eq.~(\ref{eq:reservoir}) we: let $t_o\to\infty$ (adiabatic approximation); assume an uncorrelated and thermal reservoir, \idest\ $\langle a_\ell^\dagger a_\ell\rangle=n_\ell$,  $\langle a_\ell a_\ell^\dagger \rangle=n_\ell+1$, with $n_\ell$ the Bose-Einstein distribution function, and all other expectation values of one or two bosonic operators being zero; and finally keep only the real parts of the {\it r.h.s.}\/ of eq.~(\ref{eq:reservoir}) which describe scattering and dephasing, thus neglecting the minor real-parts of energy renormalizations due to environment coupling. For the sake of brevity we omit the rather lengthy but straightforward calculation, and simply state our final result which combines the coherent and incoherent time dynamics of eqs.~(\ref{eq:coherent}) and (\ref{eq:reservoir}):

\begin{equation}\label{eq:master}
  \dot{\bm\rho}=-i(\bm h_{\rm eff}\bm\rho-\bm\rho\bm h_{\rm eff}^\dagger)
  +\J\bm\rho.
\end{equation}

\noindent Here, $\bm h_{\rm eff}=\bm h_o+\bm h_{op}-i\bm\Gamma$, $(\J\rho)_{xx'}=\sum_{\bar x\bar x'} \J_{xx',\bar x\bar x'}\bm\rho_{\bar x\bar x'}$, and:

\begin{eqnletter}\label{eq:incoherent}
  \bm\Gamma_{xx'}&=&\pi\sum_{\bar x\ell} 
  g_{x\bar x}^\ell g_{x'\bar x}^{\ell\;*} N_\ell(E_{x'}-E_{\bar x})\\
  \J_{xx',\bar x\bar x'} &=&\pi\sum_{\ell} 
  g_{x\bar x}^\ell g_{x'\bar x'}^{\ell\;^*}\left(
  N_\ell(E_{\bar x}-E_x)+N_\ell(E_{\bar x'}-E_{x'})\right),
\end{eqnletter}

\noindent with $N_\ell(\Omega)=(n_\ell+1)\delta(\omega_\ell-\Omega)+ n_\ell\delta(\omega_\ell+\Omega)$; note that in deriving eq.~(\ref{eq:incoherent}) we have assumed $g_{xx'}^\ell=g_{x'x}^{\ell\;*}$. Let us briefly comment on the physical meaning of the various contributions: in eq.~(\ref{eq:master}) the first term on the {\it r.h.s.}\/ describes a non-unitary time evolution, where $\bm\Gamma$ accounts for relaxation and dephasing due to environment coupling; in the spirit of Boltzmann's equation, these processes can be described as {\em generalized out-scatterings};\/ in contrast, $\J$ accounts for {\em in-scatterings}\/ which guarantee that the trace of $\bm\rho$ is preserved at all times. Inspection of eq.~(\ref{eq:incoherent}) reveals that these scattering contributions resemble Fermi's golden rule, and consist of emission and absorption processes associated to terms proportional to $n_\ell+1$ and $n_\ell$, respectively.

In the implementation of eqs.~(\ref{eq:master}) and (\ref{eq:incoherent}) it turns out to be convenient to introduce a few minor simplifications: firstly, we split off from the optical dipole-matrix elements the dipole moment $\mu_o$ of the bulk semiconductor and use the usual optical selection rules that light with circular polarization $\sigma_\pm$ creates electron-hole pairs with well-defined spin orientations \cite{haug:93,hohenester.pss-b:00}; hence, $\bm M_{x'x}=\sum_\sigma \mu_o\hat{\bm e}_\sigma \bar M_{x'x}^\sigma$, with $\bar M_{x'x}^\sigma$ a quantity of the order of unity; secondly, for spontaneous photon emission eq.~(\ref{eq:incoherent}) becomes:

\begin{eqnletter}\label{eq:gamma}
  \bm\Gamma_{xx'}&=&\frac {\gamma_o} 2 \sum_{\bar x\sigma}
  \bar M_{x\bar x}^\sigma \bar M_{x'\bar x}^{\sigma\;*}\\
  \J_{xx',\bar x\bar x'} &=& \gamma_o \sum_\sigma
  \bar M_{x\bar x}^\sigma \bar M_{x'\bar x'}^{\sigma\;*},
\end{eqnletter}

\noindent with $\gamma_o$ a constant of the order of the radiative decay time, \idest\ a few nanoseconds; although the value of $\gamma_o$ is fully determined by the material parameters of the bulk semiconductor, further below we shall treat $\gamma_o$ as an adjustable parameter which allows to control the strength of environment interactions.

\section{Coherent population transfer}\label{sec:stirap}

Let us first consider the $\Lambda$-type level scheme depicted in fig.~\ref{fig:scheme}{\bf a}: it consists of two long-lived states $|1\rangle$ and $|2\rangle$ which are optically connected through a third short-lived state $|3\rangle$. In ref.~\cite{hohenester.apl:00} we showed that such a scheme corresponds to the situation where a tunnel-coupled double dot is populated by one surplus hole; charging is achieved by placing the dots into a field-effect structure and applying a static electric field $F$; in addition, $F$ leads to the localization of the hole wavefunction in one of the two dots (states $|1\rangle$ and $|2\rangle$) and to an energy splitting $\delta$ between $E_1$ and $E_2$; finally, $|3\rangle$ is associated to the charged-exciton state consisting of two holes and one electron, which, because of its lighter mass, extends over both dots and thus allows optical coupling between states $|1\rangle$ and $|2\rangle$.

\begin{figure}
\centerline{
\includegraphics[width=0.48\columnwidth]{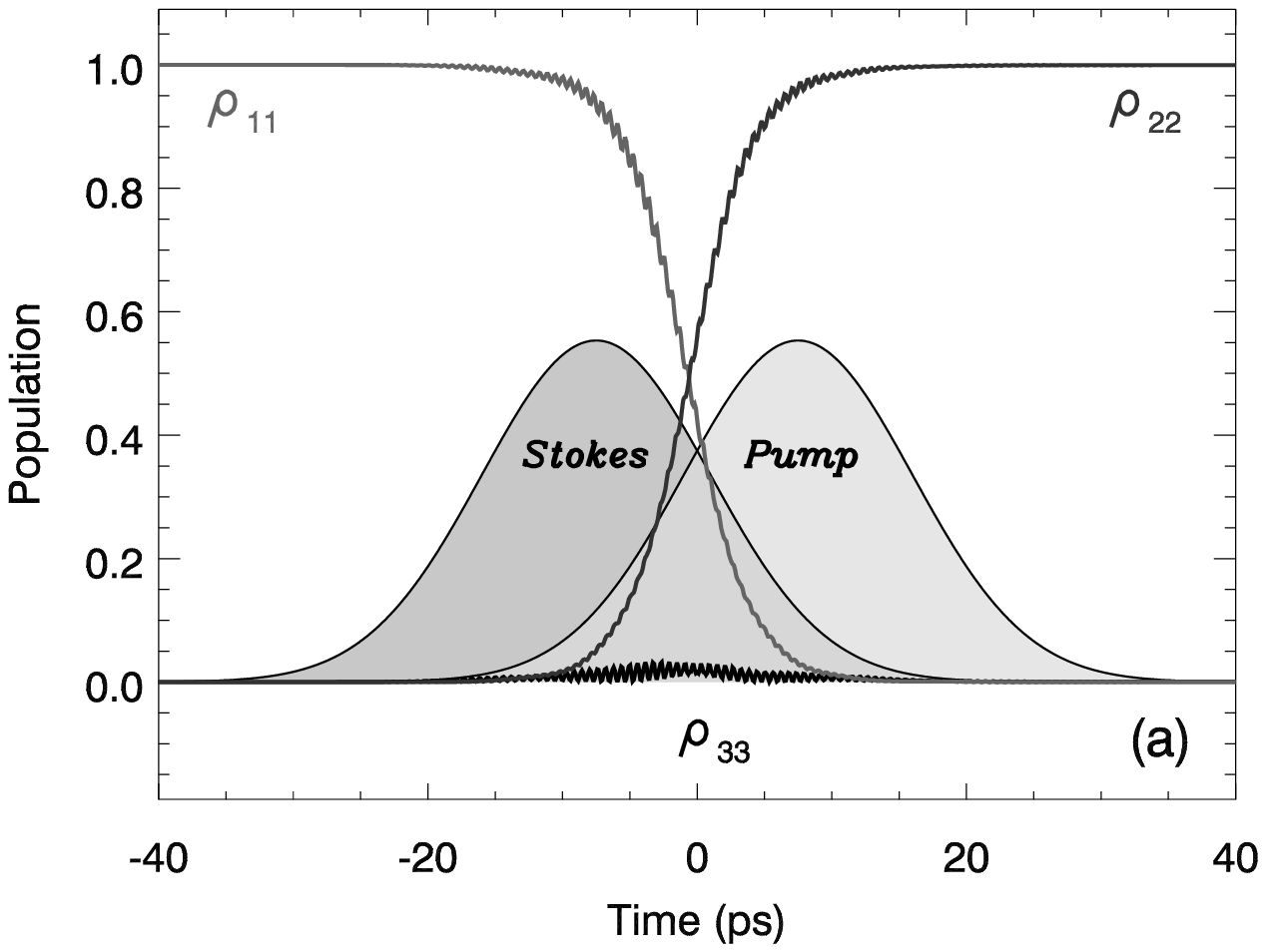}
\includegraphics[width=0.48\columnwidth]{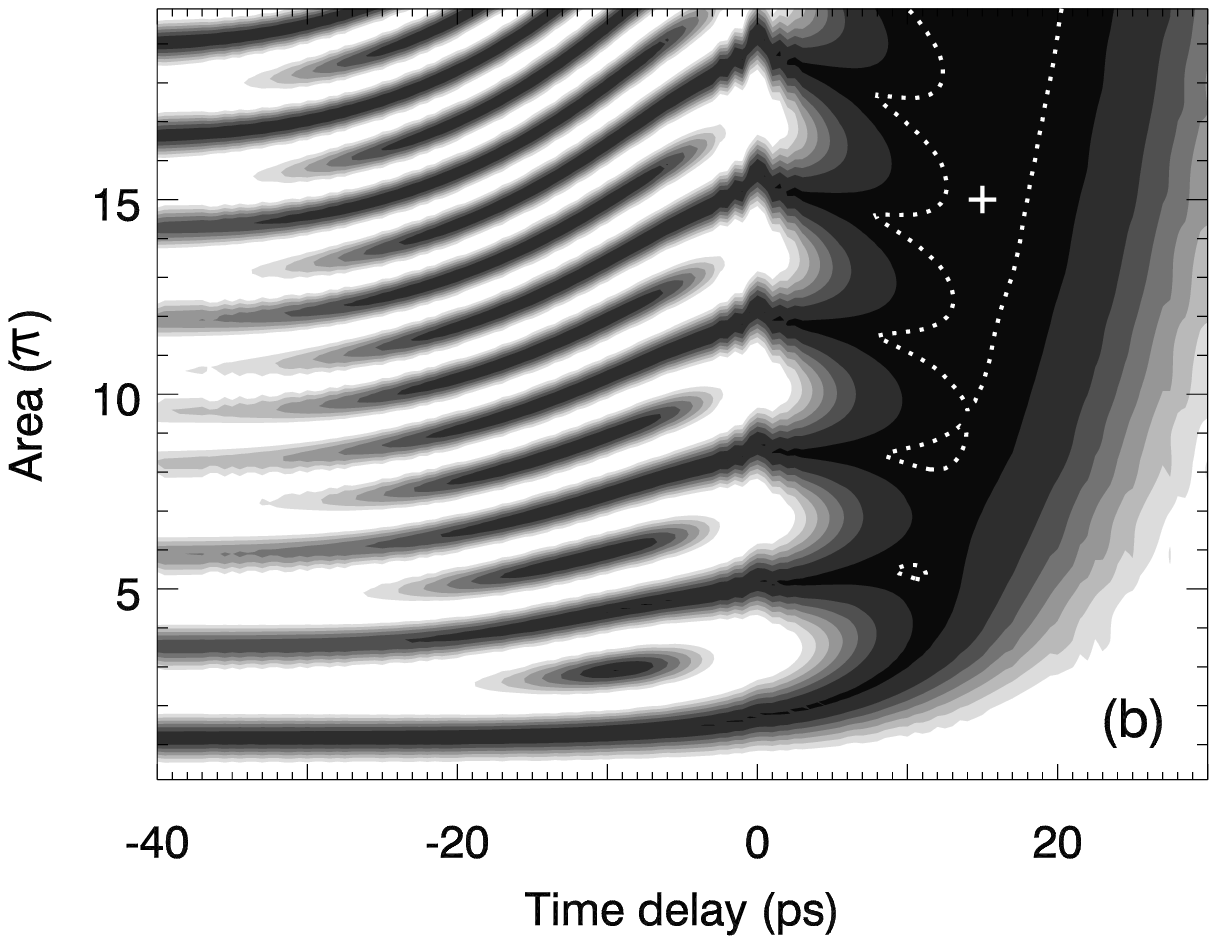}}

\caption{Simulations of coherent population transfers in coupled dots: {\bf (a)} transients of the populations $\bm\rho_{11}$, $\bm\rho_{22}$, and $\bm\rho_{33}$ ({\it cf}.\/ fig.~\ref{fig:scheme}); {\bf (b)} contour plot of final population $\rho_{22}$ as a function time delay between Stokes and pump pulse and of pulse areas ($\E_s=\E_p$); white corresponds to values below $0.1$, black to values above $0.9$; the dashed line gives the contour of $\bm\rho_{22}\ge 0.999$ and the cross indicates the values used in fig.~2{\bf a}. In our simulations we use \cite{hohenester.apl:00}: $\delta=7.1$ meV, $\gamma_o^{-1}=100$ ps, $\bar M_{13}=0.90$, $\bar M_{23}=0.79$, and Gaussian envelopes for the Stokes and pump pulse with a 20 ps full width of half maximum.}\label{fig:stirap}

\end{figure}

Quite generally, for the level scheme of fig.~\ref{fig:scheme}{\bf a} and assuming that the system is initially prepared in state $|1\rangle$, in the following we shall ask the question: what is the most efficient way to bring the system from $|1\rangle$ to $|2\rangle$? Suppose that the frequencies of two laser pulses are tuned to the 1--3 and 2--3 transitions, respectively, where, for reasons to become clear in a moment, we shall refer to the pulses as {\em pump}\/ and {\em Stokes}.\/ Since direct optical transitions between $|1\rangle$ and $|2\rangle$ are forbidden we have to use $|3\rangle$ as an auxiliary state; however, intermediate population of $|3\rangle$ introduces losses through environment coupling, \idest\ spontaneous photon emissions. Thus, which sequence of laser pulses minimizes the population of level $|3\rangle$? The answer is given in fig.~\ref{fig:stirap}{\bf b}, which shows results of our simulations for different time delays between Stokes and pump pulse and for different pulse areas $A\equiv\int_{-\infty}^\infty d\bar t\;\mu_o\E_o(\bar t)$: black (white) areas correspond to successful (no) population transfer. In the case of the ``intuitive'' ordering of laser pulses where the pump pulse excites the system {\em before}\/ the Stokes pulse, \idest\ negative time delays in fig.~2{\bf b}, one observes enhanced population transfer for odd multiples of $\pi$ of $A$, associated to processes where the pump pulse first excites the system from $|1\rangle$ to $|3\rangle$, and the subsequent Stokes pulse brings the system from $|3\rangle$ to $|2\rangle$.

However, the large black area at positive time delays in fig.~\ref{fig:stirap}{\bf b} suggests that there is a more efficient way for a population transfer: here, the two pulses are applied in the ``counter-intuitive'' order, \idest\ the pump pulse is turned on {\em after}\/ the Stokes pulse. Because of the resemblance of this scheme with a Raman-type process it has become convenient to introduce the expression of a {\em Stokes} pulse, and the whole process has been given the name {\em stimulated Raman adiabatic passage}\/ (\stirap) \cite{bergmann:98}. As will be discussed below, \stirap\ is a process which fully exploits the quantum coherence introduced by the intense laser fields: in presence of the Stokes pulse the dot-states become renormalized, and these renormalized states are used by the pump pulse for a robust and high-fidelity population transfer. While fig.~\ref{fig:stirap} presents results based on eqs.~(\ref{eq:master}) and (\ref{eq:gamma}), in our following discussion we shall introduce a few simplifying assumptions: first, we assume that the pump pulse {\em only}\/ affects the 1--3 transition and the Stokes pulse {\em only}\/ the 2--3 one, an assumption certainly valid for Rabi frequencies much smaller than $\delta$. Hence, the time evolution is governed by the effective Hamiltonian

\begin{equation}\label{eq:stirap}
  \bm h_{\rm eff}=-\frac 1 2\left(
    \begin{array}{ccc}
      0 & 0 & \Omega_p \\
      0 & 0 & \Omega_s \\
      \Omega_p & \Omega_s & i\gamma_o\\
    \end{array}\right),
\end{equation}

\noindent with $\Omega_p$ and $\Omega_s$ the Rabi frequencies for the pump and Stokes pulse, respectively. If $\Omega_p$ and $\Omega_s$ have a sufficiently slow time dependence, as will be specified in more detail further below, at each instant of time we can characterize the system by the eigenvalues and vectors of eq.~(\ref{eq:stirap}); straightforward algebra yields for the eigenvalues $\varpi_o=0$ and $\varpi_\pm\cong \frac 1 2(\Omega_{\rm eff}-i\frac{\gamma_o} 2)$, with $\Omega_{\rm eff}^2=\Omega_p^2+\Omega_s^2$ and for $\gamma_o\ll \Omega_s,\Omega_p$. Most importantly, eigenvalue $\varpi_o$ has no imaginary part and consequently does not suffer radiative losses. Indeed, introducing the time-dependent angle $\theta$ {\it viz.}\/ $\tan\theta=\frac{\Omega_p}{\Omega_s}$ we observe that the eigenvector

\begin{equation}\label{eq:trapped}
  |a_o\rangle=\cos\theta|1\rangle-\sin\theta|2\rangle
\end{equation}

\noindent has {\em no}\/ component of the ``leaky'' state $|3\rangle$, in contrast to the eigenvectors $|a_\pm\rangle$ which are composed of all three states $|1\rangle$, $|2\rangle$, and $|3\rangle$. Thus, in $|a_o\rangle$ the amplitudes of the 1--3 and 2--3 transitions interfere destructively such that the state is completely stable against absorption and emission from the radiation fields; for that reason, state $|a_o\rangle$ has been given the name {\em trapped state}.\/ The \stirap\ process uses state $|a_o\rangle$ as a vehicle in order to transfer population between states $|1\rangle$ and $|2\rangle$. Coherent population transfer is achieved by using overlap in time between the two laser pulses (see fig.~\ref{fig:stirap}{\bf a}): initially, the system is prepared in state $|1\rangle$. When the first laser pulse (Stokes) is smoothly turned on, the double-dot system is excited at frequency $\omega_s$. Apparently, at this frequency no transitions can be induced; what the pulse does, however, is to align the time-dependent state vector $|\Psi\rangle_t$ with $|a_o\rangle_t=|1\rangle$ (since $\theta=0$ in the sector $\Omega_s\neq 0$, $\Omega_p=0$), and to split the degeneracy of the eigenvalues $\varpi_o$, $\varpi_\pm$. Thus, if the second laser pulse (pump) is smoothly turned on ---such that throughout $\Omega_{\rm eff}(t)$ remains large enough to avoid non-adiabatic transitions between $|a_o\rangle_t$ and $|a_\pm\rangle_t$--- all population is transferred between states $|1\rangle$ and $|2\rangle$ within an adiabatic process where $|\Psi\rangle_t$ directly follows the time-dependent trapped state $|a_o\rangle_t$.

\stirap\ is a process important for a number of reasons. Firstly, it is a prototypical example of how intense laser fields can cause drastic renormalizations of carrier states; quite generally, these ``dressed'' states exhibit novel features in case of quantum interference, \idest\ if three or more states are optically coupled. Secondly, pulse-shaping techniques in quantum dots have recently attracted increasing interest in view of possible quantum computation applications \cite{troiani.prb:00,chen:01b,pazy:01}, aiming at an all-optical control of carrier states; in this respect, \stirap\ might be of some importance because of its robustness and its high fidelity. More specifically, from fig.~\ref{fig:stirap}{\bf b} it becomes apparent that population transfer works successfully within a relatively large parameter regime---in contrast to the ``intuitive'' order of pulses, where a detailed knowledge of the dipole matrix elements and a precise control over the laser pulses is required. Thus, \stirap\ is a robust scheme which only relies on sufficiently smooth and strong laser pulses; we note here in passing that the population transfer is governed by the temporal overlap of the pulses, thus also allowing for the creation of superposition states through appropriate tailoring of the pulse envelopes \cite{marte:91,unanyan:01,kis:02}. Finally, \stirap\ processes minimize losses through environment coupling and therefore work with high fidelity; one can envision that such high-quality performance might be required for quantum computation implementations where information is encoded in spin degrees of freedom and optical control is used for quantum gates, whereby scattering processes during the gating might become the primary source of dephasing.

\section{Self-induced transparency}\label{sec:sit}

In our second example we discuss the propagation of an intense laser beam in a sample of inhomogenously broadened quantum dots: as will be shown in the following, above a given threshold the quantum coherence introduced by the laser leads to a drastic modification of the non-linear optical properties, and stable pulse propagation without significant losses becomes possible. Throughout we shall assume that the laser frequency is tuned to the exciton groundstate, and describe the carrier states in terms of the level scheme depicted in fig.~\ref{fig:scheme}{\bf b}. The description of an ensemble of quantum dots is achieved by simulating a large number of dots, \idest\ typically hundred, and computing the material response as an appropriate ensemble average; most importantly, the macroscopic interband polarization $\bm\P$ then reads:

\begin{equation}\label{eq:ensemble}
  \bm\P=\mathcal{N}\int g(\Delta)d\Delta\sum_{xx'} 
  \bm M_{x'x}\bm\rho_{xx'}(\Delta),
\end{equation}

\noindent with $\mathcal{N}$ the dot density, $\Delta$ the exciton detuning, and $g(\Delta)$ the normalized distribution function accounting for inhomogeneous broadening which is characterized by the full-width of half maximum $\delta^*$. In what follows, it is crucial to note that $\bm\P$ is not only driven by the laser field but, on its part, also acts as source for the electric field. To account for this back-action we additionally compute ---within the rotating-wave and slowly-varying envelope approximations \cite{mandel:95,panzarini.prb:02}--- the time evolution of $\bm\E_o$ through:

\begin{equation}\label{eq:maxwell}
  \left(\partial_z+\frac n c\partial_t\right)\bm\E_o(z,t)\cong
  \frac{2\pi\omega_o}{nc}\mbox{Im}\bm\P(z,t),
\end{equation}

\noindent with $n$ the semiconductor refraction index, $c$ the speed of light, and assuming pulse propagation in the positive $z$-direction. Computationally, we solve the coupled set of eqs.~(\ref{eq:master},\ref{eq:ensemble},\ref{eq:maxwell}) on a sufficiently dense real-space grid $z_i$, \idest\ typically thousand points, where $\bm\P(z,t)$ is calculated for each $z_i$ from eq.~(\ref{eq:ensemble}).

\begin{figure}
\centerline{
\includegraphics[height=2.8in]{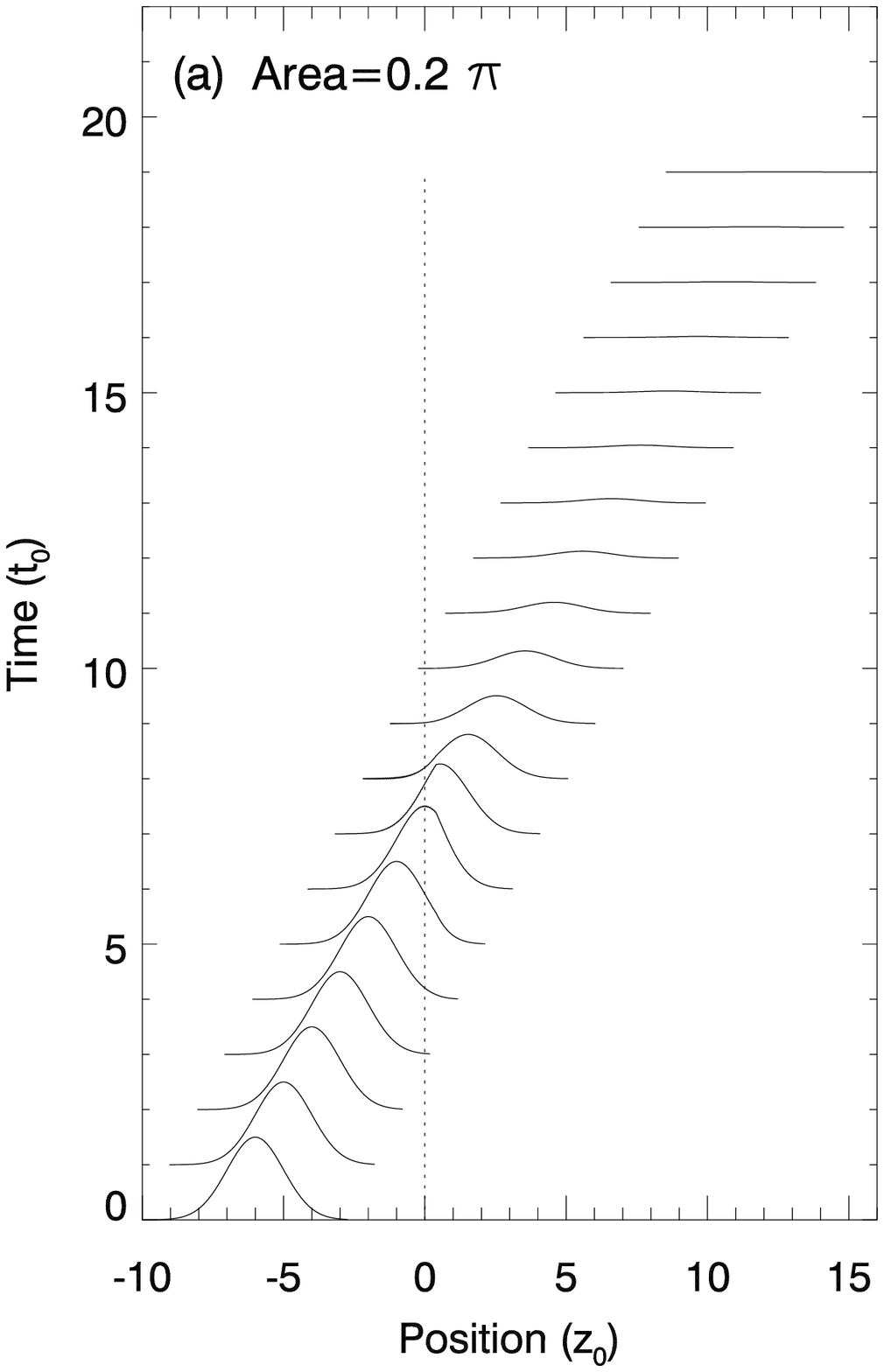}
\includegraphics[height=2.8in]{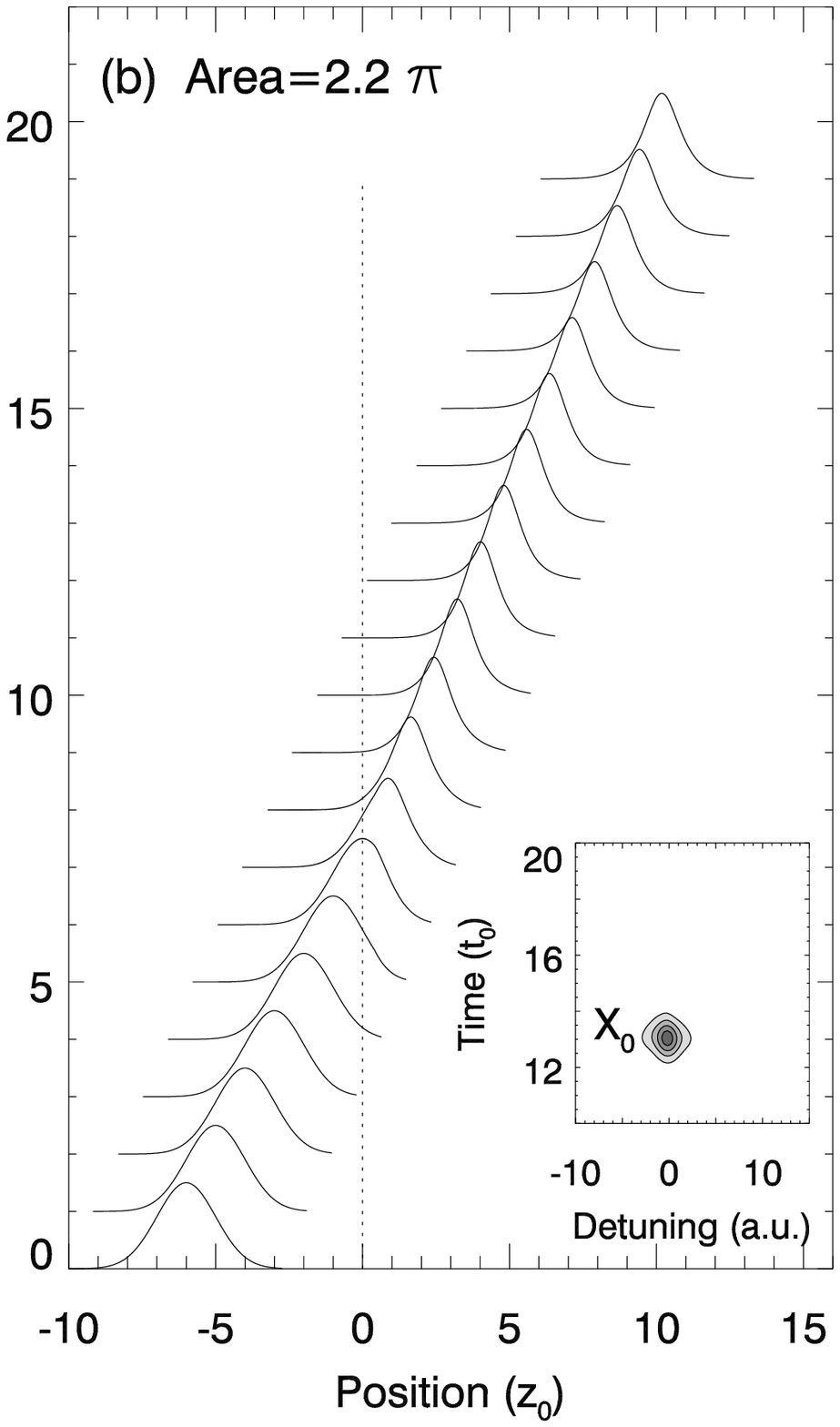}
\includegraphics[height=2.8in]{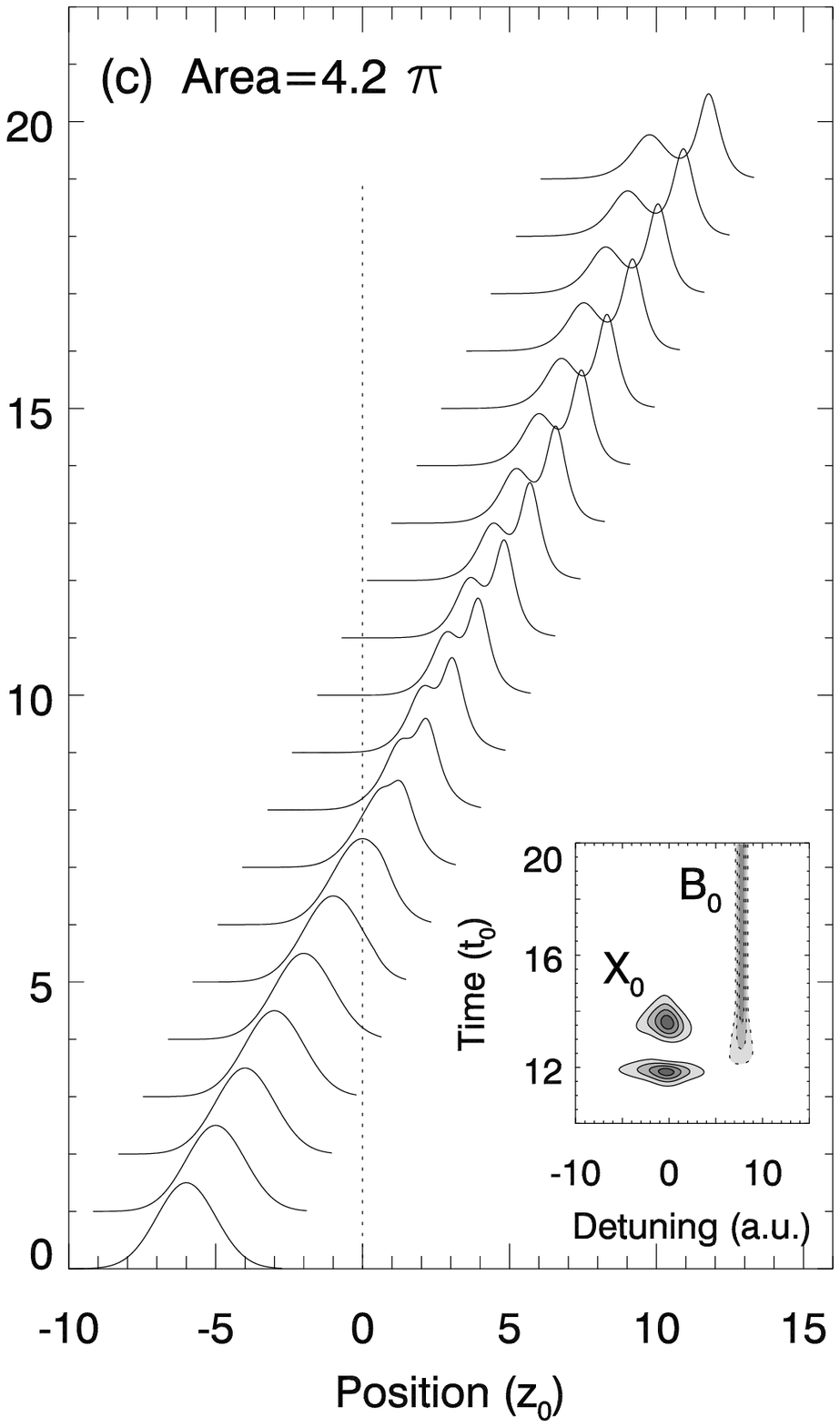}
}

\caption{Results of our simulations (\idest, $\mathcal{E}_o(z,t)$ in arbitrary units) of pulse propagation in a sample of inhomogenously broadened quantum dots and for different pulse areas; we assume linear polarization and a setup where the pulse enters from a dot-free region (negative $z$-values) into the dot region. The insets show contour plots of the time evolution of $\bm\rho(\Delta)$ for $z=5$, with $\Delta$ measured in units of $\hbar/t_o$; we use a biexciton binding energy of 16 a.u. and $\bar M=1$ for all optically allowed transitions.}\label{fig:sit}

\end{figure}

Before discussing the results of our simulations, fig.~\ref{fig:sit}, let us briefly address the basic characteristics expected for such a pulse propagation. Quite generally, we shall assume that:

\begin{equation}\label{eq:sit}
  \gamma_o\ll \tau_o^{-1} \ll\delta^*,
\end{equation} 

\noindent with $\tau_o$ the temporal width of the laser pulse; in other words, the homogeneous level broadening should be much smaller than the spectral width of the laser pulse, and both broadenings should be much smaller than the inhomogeneous broadening $\delta^*$. Apparently, for typical dot and laser parameters $\gamma_o^{-1}\sim 1$ ns, $\tau_o\sim 1$--10 ps, and $\delta^*\sim 10$ meV, eq.~(\ref{eq:sit}) is easily fulfilled. As regarding the general trends of pulse propagation, we first concentrate on the case of a weak laser pulse entering the dot region: apparently, the laser will excite excitons and hereby suffer attenuation; a more detailed analysis reveals an exponential damping (Beer's law of linear absorption) with $z_o=nc/ (2\pi^2\mathcal{N}\omega_o\mu_o^2g(0))$ providing a characteristic length scale \cite{mandel:95,mccall:67,mccall:69}; in ref.~\cite{panzarini.prb:02} we estimate a value $z_o\sim 0.15$ mm for typical InGaAs dot samples. Because of the weak dephasing the laser-induced coherence keeps stored in the material even after attenuation of the laser pulse. Indeed, in the pioneering work of {\sc mc call} and {\sc hahn} \cite{mccall:67,mccall:69} the authors showed for a proto-typical two-level system that above a given threshold this stored energy can again be fully extracted from the material and given back to the laser pulse. To come to this conclusion the authors made two important observations: first, there exists a specific pulse shape, \idest\ hyperbolic secant, for which all two-level systems are driven through a sequence of exited states back to their groundstates irrespective of the detuning $\Delta$, provided that the pulse area is a multiple of $2\pi$; second, a laser pulse of arbitrary shape entering a region of inhomogenously broadened two-level systems achieves in a self-modulation-like process this hyperbolic secant shape. Consequently, it can propagate without suffering significant losses such that at each instant of time the pulse gives and receives the same amount of energy from the material; thus, this striking phenomenon has been given the name {\em self-induced transparency}\/ (\sit).

Indeed, such behavior is observed in fig.~\ref{fig:sit} (position measured in units of $z_o$ and time in units of $t_o=nz_o/c$) which shows results of simulations using the level scheme of fig.~\ref{fig:scheme}{\bf b}: for small field strengths, fig.~\ref{fig:sit}{\bf a}, the pulse becomes attenuated quickly; however, if the pulse area exceeds a certain value, fig.~\ref{fig:sit}{\bf b}, self modulation occurs and the pulse propagates without suffering significant losses; the inset of fig.~\ref{fig:sit}{\bf b} shows a contour plot of the exciton population as a function of $\Delta$ and time, where one clearly observes the above-mentioned excitation and de-excitation of exciton states; finally, at the highest pulse area, fig.~\ref{fig:sit}{\bf c}, we observe pulse breakup \cite{mandel:95,mccall:67,mccall:69}; the inset shows a $4\pi$-rotation of the exciton states and an additional population of the biexciton channel. As apparent from the figure, this biexciton population, which is attributed to the high field strengths corresponding to Rabi frequencies comparable to the biexciton binding, does not spoil the general pulse propagation properties. However, closer inspection reveals that such strong laser pulses create a macroscopic entanglement of exciton states $X_o^\pm$, as will be discussed in more detail elsewhere.

\section{Conclusions and outlook}\label{sec:conclusion}

In conclusion, we have developed a general theoretical framework suited for the description of coherent and incoherent dynamics in optically excited semiconductor quantum dots. This model has been used for the study of two prototypical situations where intense laser fields lead to pronounced renormalizations of carrier states: first, it has been shown how two intense laser pulses can be exploited for a robust and high-fidelity population transfer of carrier states in coupled quantum dots; second, we have discussed how intense laser pulses can propagate in a sample of inhomogenously broadened quantum dots without suffering significant losses. In a sense, these two examples, as well as the numerous recent reports on related subjects, clearly demonstrate that many surprising effects which hitherto have only been seen in atomic systems have now also become observable in artificial atoms and molecules.

Quite generally, the success of {\em Quantum Optics}\/ steams from the fact that atoms have been studied for many decades, with all fine details of the level structure now being well understood, and that atoms provide both short-lived states, with lifetimes of the order of nanoseconds, as well as states extremely well protected from their environment, with lifetimes of the order of seconds. As concerning artificial atoms and molecules, the situation is less clear: on the one hand, there now exists some consensus regarding the basic characteristics of few-particle states in single dots, whereas many issues regarding excited states or properties of artificial molecules still remain to be explored; on the other hand, the more intimate coupling of quantum dot states to their solid-state environment does not seem to permit the same lifetime diversity as compared to atoms. Yet, long-lived excitations, {\it e.g.}\/ spin degrees of freedom, and possible manipulation channels still remain to be identified, and will ultimately make possible or not the solid-state implementation of quantum computation or related schemes in this class of material.

On a more fundamental level, the success of quantum optics in quantum dots will strongly depend on the basic differences and on the added value of this class of material in comparison with the high standards of quantum optics in atomic systems. In this respect, bi- and multi-excitons, \idest\ excitations consisting of several electron-hole pairs, have recently been demonstrated in the context of single-photon sources as the first of such a clear-cut peculiarity of quantum dots. Thus, risklessly we can say that the whole field of quantum optics in quantum dots is still in its infancy and that many exciting prospects have not even been foreseen today.

\acknowledgments

Our deepest thanks go to Giovanna Panzarini, who introduced us to most of the ideas that are presented in this paper and who shared with us her physical intuition and her joy for the beauty of physics. The main pleasure in collaborating with Giovanna was to see her studying literature, discovering dozens of new things, and discussing them with us. She was a hard-working woman who was moving fast from one topic to the other, sometimes a little bit too fast, as if she already knew about the little time that was left. Personally, Giovanna kept distance. Although being truly and sympathetically interested in the news of others, her personal life often remained protected. We all miss her so much. Her passion for science, her brilliant and rigorous comments, her smile and her attentive and passionate presence in the group will always be in our memory. We will remember the picture of Giovanna entering the room with a smile on her face, saying that she just discovered a beautiful, little effect.


\begin{thebibliography}{10}

\bibitem{mandel:95}
L. Mandel and E. Wolf, {\em Optical coherence and quantum optics} 
  (Cambridge University Press, Cambridge, 1995).

\bibitem{scully:97}
M.~O. Scully and M.~S. Zubairy, {\em Quantum optics} (Cambridge University
  Press, Cambridge, UK, 1997).

\bibitem{nature:02}
See, e.g., S. Chu, Nature, {\bf 206} (2002) 206, and references therein.

\bibitem{bouwmeester:00}
{\em The Physics of Quantum Information}, D. Bouwmeester, A. Ekert, and A.
  Zeilinger, eds., (Springer, Berlin, 2000).

\bibitem{woggon:97}
U. Woggon, {\em Optical properties of semiconductor quantum dots} (Springer,
  Berlin, 1997).

\bibitem{hawrylak:98}
L. Jacak, P. Hawrylak, and A. Wojs, {\em Quantum dots} (Springer, Berlin,
  1998).

\bibitem{bimberg:98}
D. Bimberg, M. Grundmann, and N. Ledentsov, {\em Quantum dot heterostructures}
  ({J}ohn {W}iley, New York, 1998).

\bibitem{zrenner:00}
A. Zrenner, J. Chem. Phys. {\bf 112} (2000) 7790.

\bibitem{simserides.prb:00}
C. Simserides, U. Hohenester, G. Goldoni, and E. Molinari, 
  Phys. Rev. B {\bf 62} (2000) 13657.

\bibitem{warburton:00}
R.~J. Warburton, C. Sch{\"a}flein, D. Haft, F. Bickel, A. Lorke, K. Karrai,
  J.~M. Garcia, W. Schoenfeld, and P.~M. Petroff, 
  Nature {\bf 405} (2000) 926.

\bibitem{hartmann.prl:00}
A. Hartmann, Y. Ducommun, E. Kapon, U. Hohenester, and E. Molinari,
  Phys. Rev. Lett. {\bf 84} (2000) 5648.

\bibitem{findeis.prb:01}
F. Findeis, M. Baier, A. Zrenner, M. Bichler, G. Abstreiter, U. Hohenester, and
  E. Molinari, 
  Phys. Rev. B {\bf 63} (2001) 121309(R).

\bibitem{bayer:00a}
M. Bayer, O. Stern, P. Hawrylak, S. Fafard, and A. Forchel, 
  Nature {\bf 405} (2000) 923.

\bibitem{hohenester.pss-b:00}
U. Hohenester and E. Molinari, 
  phys. stat. sol. (b) {\bf 221} (2000) 19.

\bibitem{verzelen:00}
O. Verzelen, R. Ferreira, and G. Bastard,
  Phys. Rev. B {\bf 62} (2000) R4809.

\bibitem{uskov:00}
A.~V. Uskov, A.~P. Jauho, B. Tromborg, J. Mork, and R. Lang, 
  Phys. Rev. Lett. {\bf 85} (2000) 1516.

\bibitem{borri:01}
P. Borri, W. Langbein, S. Schneider, U. Woggon, R.~L. Sellin, D. Ouyang, and D.
  Bimberg, Phys. Rev. Lett. {\bf 87} (2001) 157401.

\bibitem{schedelbeck:97}
G. Schedelbeck, W. Wegscheider, M. Bichler, and G. Abstreiter, 
  Science {\bf 278} (1997) 1792.

\bibitem{bayer:01a}
M. Bayer, P. Hawrylak, K. Hinzer, S. Fafard, M. Korkusinski, R. Wasilewski, O.
  Stern, and A. Forchel, Science {\bf 291} (2001) 451.

\bibitem{troiani.prb:02}
F. Troiani, U. Hohenester, and E. Molinari,
  Phys. Rev. B {\bf 65} (2002) 161301(R).

\bibitem{troiani.prb:00}
F. Troiani, U. Hohenester, and E. Molinari, 
  Phys. Rev. B {\bf 62} (2000) R2263.

\bibitem{biolatti:00}
E. Biolatti, R.~C. Iotti, P. Zanardi, and F. Rossi,
  Phys. Rev. Lett. {\bf 85} (2000) 5647.

\bibitem{haug:93}
H. Haug and S.~W. Koch, {\em Quantum theory of the optical and electronic
  properties of semiconductors} (World Scientific, Singapore, 1993).

\bibitem{rontani.ssc:01}
M. Rontani, F. Troiani, U. Hohenester, and E. Molinari,
  Solid State Commun. {\bf 119} (2001) 309.

\bibitem{leggett:87}
A.~J. Leggett, S. Chakravarty, A.~T. Dorsey, M.~P.~A. Fisher, A. Garg, and W.
  Zwerger, Rev. Mod. Phys. {\bf 59} (1987) 1.

\bibitem{hohenester.ssc:01}
U. Hohenester, Solid State Commun. {\bf 118} (2001) 151.

\bibitem{walls:95}
D.~F. Walls and G.~J. Millburn, {\em Quantum Optics} (Springer, Berlin, 1995).

\bibitem{hohenester.apl:00}
U. Hohenester, F. Troiani, E. Molinari, G. Panzarini, and C. Macchiavello,
  Appl. Phys. Lett. {\bf 77} (2000) 1864.

\bibitem{bergmann:98}
K. Bergmann, H. Theuer, and B.~W. Shore, 
  Rev. Mod. Phys. {\bf 70} (1998) 1003.

\bibitem{chen:01b}
P. Chen, C. Piermarocchi, and L.~J. Sham,
  Phys. Rev. Lett. {\bf 87} (2001) 067401.

\bibitem{pazy:01}
E. Pazy, I. {D'Amico}, P. Zanardi, and F. Rossi, 
  Phys. Rev. B {\bf 64} (2001) 195320.

\bibitem{marte:91}
P. Marte, P. Zoller, and J.~L. Hall, 
  Phys. Rev. A {\bf 44} (1991) R4118.

\bibitem{unanyan:01}
R.~G. Unanyan, N.~V. Vitanov, and K. Bergmann,
  Phys. Rev. Lett. {\bf 87} (2001) 137902.

\bibitem{kis:02}
Z. Kis and F. Renzoni,
  Phys. Rev. A {\bf 65} (2002) 032318.

\bibitem{panzarini.prb:02}
G. Panzarini, U. Hohenester, and E. Molinari, 
  Phys. Rev. B {\bf 65} (2002) 165322.

\bibitem{mccall:67}
S.~L. McCall and E.~L. Hahn, 
  Phys. Rev. Lett. {\bf 18} (1967) 908.

\bibitem{mccall:69}
S.~L. McCall and E.~L. Hahn, 
  Phys. Rev. {\bf 183} (1969) 457.
  
\end{thebibliography}
\end{document}